\documentclass[letterpaper,twocolumn,10pt]{article}
\usepackage{usenix-2020-09}

\usepackage[utf8]{inputenc}
\usepackage{amsmath,amssymb}
\usepackage{booktabs}
\usepackage{flushend}
\usepackage{caption}
\usepackage{tabularx}
\usepackage{multirow}
\usepackage{xcolor}
\usepackage{hyperref}
\usepackage{url}
\usepackage{xspace}
\usepackage{enumitem}
\usepackage{tikz}
\usetikzlibrary{arrows.meta,positioning,fit}

\newcommand{\Plan}{\textsc{Plan}\xspace}
\newcommand{\GiB}{\,GiB\xspace}
\newcommand{\MiB}{\,MiB\xspace}
\newcommand{\para}[1]{\smallskip\noindent\textbf{#1.}}
\newcommand{\fullhash}[8]{\texttt{#1#2#3#4}\allowbreak\texttt{#5#6#7#8}}
\newcommand{\commitid}[5]{\texttt{#1#2}\allowbreak\texttt{#3#4#5}}
\newcommand{\claimref}[1]{\hyperlink{claim:#1}{\textsuperscript{\textsf{#1}}}}
\hypersetup{colorlinks=true,allcolors=black}
\setlist[itemize]{leftmargin=*,nosep}
\begin{document}

\title{Memory-Sovereign Inference:\\[-0.15em]Output-Exact Execution Beyond Full Residency}
\author{Lukas Stepanek\\\texttt{luki.step@proton.me}}
\date{Technical report}
\maketitle

\begin{abstract}
Storage-backed inference is easy to demonstrate and easy to overclaim.  A low
process resident set does not bound charged page cache, a process-local device
reading does not establish whole-board use, and successful generation does not
show that asynchronously reused buffers held the intended bytes.  We present
a reference architecture and falsifiable execution-evidence certificate.  A
certificate records an immutable representation, separates
semantic workload demand from scheduler requests and cumulative traffic, names
a per-tier resource-authority vector and exactness horizon, and records
implementation-independent reuse invariants with the transitions actually
tested.

At one Qwen3-Next system identity, the complete native tensor representation is
45.08\GiB.  The stock router selects every one of the 48$\times$512 managed
layer--expert objects during the 32K prefill.  Their duplicate-free,
overlap-free canonical-extent union yields a conservative 43.59375\GiB semantic-
demand lower bound.  Both quantities exceed the declared 34\GiB full-residency
envelope; the demand lower bound also exceeds the 11\GiB host-hard plus physical
24\GiB-device envelope by 8.59375\GiB\claimref{C1}.  The LRU64 execution
remains within its host-hard/GPU-audited resource contract.  Against one
prespecified zero-cache oracle, all 64 token IDs, every byte of all 64 complete
151,936-float logit rows, all 3,408 recorded route events, response bytes, and
the seven-field consumer and two-entry destination identities are exact.  We
call this relation full-output-horizon
exactness; recurrent-state and upstream-runtime equality are not claimed.

In a matched source campaign, buffered P completes exactly but reaches the
11\GiB host ceiling and records 33,481 \texttt{memory.max} events.  Blocking
one-window direct D and the complete eight-window asynchronous F component are
exact with positive margin and zero limit events.  Across six counterbalanced
pairs, F completes in 32.3\% of D's wall time (3.10$\times$ reciprocal); all
F/D ratios lie in [0.3211, 0.3283], at identical physical source bytes per
output\claimref{C2}.  The contrast jointly changes queue depth, overlap, and
lifecycle implementation.  In a separate package of later fault-instrumented
executables, all fourteen
prespecified deterministic control/fault cells satisfy their validators across
Qwen3-Next and Gemma 4 adapters\claimref{C3}.  They support fail-closed behavior
for named transitions in those binaries, not a reliability rate, race-freedom
claim, or fault attribution to the earlier 32K+64 executable.  The principal
experiment remains one fixed model, workload, runtime, device, and 64-output
horizon.
\end{abstract}

\section{Introduction}
\label{sec:intro}

CPU offload, memory mapping, SSD streaming, expert caching, and speculative
generation all trade capacity for traffic or latency
\cite{flexgen,deepspeedinference,deepnvme,llminflash,moeinfinity,flexinfer,specoffload}.
The difficult question is what a successful run proves.  A process RSS limit
does not bound the page cache.  A device allocation reported by one process
does not include the rest of the board.  A cache hit is invalid if a slot is
recycled before its last consumer.  A final token match can hide stale reads
that did not affect one prompt.  Consequently, a model can appear to fit an
operator's memory budget without establishing either resource admissibility or
execution integrity.

We call the stronger property \emph{memory sovereignty}.  A memory-sovereignty
evidence record states what bytes exist and are semantically demanded, which
authority governs each resource tier, what equivalence relation holds for which
horizon, and which reuse invariants were tested.  A claim class then states what
that record qualifies.  Host enforcement and GPU accounting need not use the
same mechanism: here, process-tree host memory is hard-limited with cgroup v2,
whereas GPU use is declared and audited from whole-board observations; the
physical board supplies a separate hard capacity ceiling.  The authority vector
is part of the result, not a presentation detail.

The architectural observation is that four concerns commonly fused into a
tensor allocation should be separated: \emph{canonical storage ownership},
\emph{physical residency}, \emph{stable execution addresses}, and
\emph{scheduling}.  The architecture realizes this separation through a shared \Plan.
The plan maps logical objects to byte-preserving source extents, fixed host
windows, reusable GPU slots, stable addresses, and generations.  Model adapters
name the objects required by a layer; the native graph, kernels, router, KV
state, and sampler remain authoritative.

The contribution is an execution-certification methodology and reference
realization for storage-backed inference, not a new offload primitive.  Direct I/O, bounded
windows, stable addresses, generations, caching, and prefetch appear across
FlexInfer, MESH, Palm, MoE-Direct, FluxMoE, and DeepNVMe
\cite{flexinfer,mesh,palm,moedirect,fluxmoe,deepnvme}.  The differentiated result
is a publication evidence package: the principal Qwen3-Next experiment executes a workload
whose semantically demanded managed weights exceed the declared and host-hard/
device-physical envelopes with full-output-horizon exactness and hard complete-
process-tree host accounting; a later matched-source experiment shows direct ownership
changing the prespecified admission outcome and the complete asynchronous
component completing in 32.3\% of a blocking direct reader's wall time
(3.10$\times$ reciprocal); and separate fault-instrumented binaries pass named
fail-closed transitions across two adapters.  The package does not collapse
these executable identities into one certificate.

This report makes four contributions:

\begin{itemize}
  \item It demonstrates full-output-horizon exact native Qwen3-Next execution
  under $E_{\mathrm{64-out}}$ when a 43.59375\GiB semantic-demand lower bound
  exceeds a declared 34\GiB full-residency envelope, with hard
  complete-process-tree cgroup-charge accounting and audited whole-board GPU
  margin\claimref{C1}.
  \item Relative to the tested buffered P path, it shows that direct source
  ownership changes the prespecified hard-margin admission outcome, and it
  shows the complete eight-window asynchronous component finishing in 32.3\% of
  a blocking one-window direct reader's wall time (3.10$\times$ reciprocal) at
  identical physical source bytes\claimref{C2}.
  \item In later fault-instrumented builds, it tests all fourteen prespecified control/fault
  cells through two real model adapters, supporting named fail-closed
  transitions at those tested identities\claimref{C3}.
  \item It defines a falsifiable, implementation-independent certificate and a
  shared cross-adapter \Plan realization, then records the temporal,
  Qwen-specific, and Gemma experiments without pooling their identities.
\end{itemize}

The report also records stopped lines.  A GPT-OSS packing optimization did not
repair its physical carrier, and a Llama-3.3 placement result passed resources
but failed its task-admission criterion.  These negatives prevent local engineering
wins from being promoted into an inference-system claim.

\section{The Memory-Sovereignty Contract}
\label{sec:contract}

\subsection{Certificate Tuple and Authority Grades}

We represent an execution-evidence record as
$\mathcal{C}=(I,B,A,E,L)$.  $I$ is an immutable inventory and provenance record;
$B$ is the boundary vector by memory tier; $A$ assigns each boundary an
authority grade; $E$ names the equivalence relation, oracle identity, and
horizon; and $L$ names the reuse invariants and transition set actually tested.
Authority grades distinguish a hard-enforced limit, a physical capacity
ceiling, an instrumented high-water mark, and a sampled audit.  A certificate
may therefore report host=hard and GPU=audited without treating the two claims
as equivalent.

The tuple is a record schema, not by itself a pass criterion.  We attach
compositional claim classes: \textsc{R} states that the complete
representation cannot be resident under $B$; \textsc{W} states the stronger
workload result that a semantic-demand lower bound cannot be resident under
$B$; $\textsc{E}[E,h]$ names an exact relation and horizon; and
$\textsc{L}[F]$ names a tested transition set.  A memory-sovereign execution
claim requires \textsc{R} or \textsc{W}, a passing resource predicate
under the disclosed authority vector, a named $E$, and clean applicable
lifecycle closure at one execution identity.  Additional fault cells may form a
separate \textsc{L} record, but do not transfer between binaries implicitly.
Classes are compared dimension by dimension; the paper defines no scalar grade
that would conflate hard enforcement, physical capacity, and sampled audit.

The inventory separates four quantities.  $T_{\mathrm{repr}}$ is the complete
canonical tensor payload under a stated manifest inclusion rule.
$U_{\mathrm{sem}}^{\mathrm{LB}}$ is a conservative lower bound on unique logical
canonical bytes selected by the authoritative model graph and consumed at least
once.  $U_{\mathrm{sched}}$ is the deduplicated union requested by a source or
cache policy.  $Q_{\mathrm{run}}$ is cumulative logical or physical traffic,
including rereads and alignment padding.  Scheduler activity and traffic are
never substituted for semantic demand.

Let $H_{\mathrm{cap}}$ be the enforced process-tree host allowance,
$G_{\mathrm{decl}}$ the declared and audited GPU allowance, $G_{\mathrm{phys}}$
the physical board capacity, and $R_{\mathrm{policy}}$ an operator-declared
reserve for non-target host state.  Let $S$ denote the canonical bytes managed
from storage.  The evaluated predicates are:

\begin{align}
S &> H_{\mathrm{cap}}-R_{\mathrm{policy}},
  \label{eq:hostnonresident}\\
T_{\mathrm{repr}} &> H_{\mathrm{cap}}+G_{\mathrm{decl}},
  \label{eq:reprnonresident}\\
U_{\mathrm{sem}}^{\mathrm{LB}} &>
  H_{\mathrm{cap}}+G_{\mathrm{phys}}.
  \label{eq:demandnonresident}
\end{align}

Equation~\ref{eq:hostnonresident} is policy-relative: $S$ cannot coexist in the
host allowance while preserving the declared reserve.  Equation
~\ref{eq:reprnonresident} makes complete representation residency infeasible
within the declared placement envelope.  Equation~\ref{eq:demandnonresident}
is a semantic-workload result against a host-hard/device-physical envelope.
Adding host and GPU bytes is a relaxation despite their operational
non-fungibility; exceeding the sum is therefore sufficient for nonresidency.

Separately, the evaluated workstation exposes 16,667,594,752 host bytes through
Linux and a physical 25,769,803,776-byte GPU board.  The Qwen3-Next complete
representation exceeds their sum by 5.56\GiB, and its demanded-expert lower
bound exceeds it by 4.07\GiB\claimref{C1}.  This is an operating-system-visible
capacity observation, not a claim about all physically installed or
firmware-reserved host DRAM.

A virtual mapping does not discharge these predicates.  Buffered I/O can move
model bytes into charged page cache while process RSS stays low.  Claim-bearing
runs therefore place the complete process tree in a cgroup v2 boundary,
disable swap, and record \texttt{memory.current}, \texttt{memory.peak}, swap,
OOM, and \texttt{memory.max} events.  Device observers sample whole-board
allocation and compute-process ownership.  GPU compliance is audited rather
than hard-partitioned: a valid cell has positive margin below its declared
allowance and no foreign compute process; the board size is used separately as
a physical ceiling.

We call the host predicate \emph{hard complete-process-tree cgroup-charge
accounting}: it covers all memory charges that cgroup v2 attributes to the
target hierarchy, including charged anonymous memory, file cache, and kernel
memory.  It is not a total-causal-host-byte claim.  Shared pages retain Linux
charge semantics; firmware, device-internal buffers, unaccounted driver
allocations, and processes outside the hierarchy are outside this predicate.
No cooperating process outside the hierarchy participates in the recorded
cells.  The principal telemetry archives the aggregate high-water mark and
events, not a per-class \texttt{memory.stat} decomposition.

\begin{table}[t]
  \centering
  \small
  \caption{Implementation-independent certificate obligations.  Generations,
  epochs, and consumer counts are this prototype's realization, not universal
  requirements.}
  \label{tab:contract}
  \begin{tabularx}{\columnwidth}{@{}p{0.17\columnwidth}X@{}}
    \toprule
    Field & Required evidence \\
    \midrule
    $I$ & Representation, semantic-demand and scheduler unions, cumulative
      traffic, inclusion rules, and immutable provenance. \\
    $B,A$ & Per-tier allowance and authority grade; process-tree/page-cache and
      whole-board attribution; margins and event counters. \\
    $E$ & Same-build oracle, compared surfaces, and exact horizon; semantic
      metrics remain separate. \\
    $L$ & No publication before movement; requested object/version identity;
      no reuse before consumers finish; reset/shutdown drain or invalidate;
      invalid completions cannot make data visible. \\
    \bottomrule
  \end{tabularx}
\end{table}

\subsection{Exactness Is Scoped by Identity and Horizon}

For the principal experiment, the named relation is
\begin{align}
E_{\mathrm{64-out}}={}&E_{\mathrm{logits}}\land E_{\mathrm{tokens}}
  \land E_{\mathrm{response}}\notag\\
 &{}\land E_{\mathrm{routes}}\land E_{\mathrm{consumer}}
  \land E_{\mathrm{destination}}.
\end{align}
Here $E_{\mathrm{logits}}$ covers every byte of all 64 complete logit rows;
the remaining terms cover the corresponding ordered tokens, response bytes,
recorded routes, consumer identity, and destination identity.
We use \emph{full-output-horizon exactness under $E_{\mathrm{64-out}}$}; the
phrase does not imply equality of every internal state.

The consumer term compares a seven-field record naming the base and adapter
commits, graph, router, expert kernel, reduction, and scheduling boundary.  The
destination term compares the two-entry model-memory placement record.  Both
are compared structurally, rather than inferred from output equality.

The strongest oracle compares complete logits and generated token IDs against
the same build, graph, tensor representation, prompt, and deterministic
sampler.  Sparse routes and recurrent state are included when exposed by the
runtime.  The term \emph{exact} applies only to the recorded scope.  A short test
does not certify a longer execution, and a prefix comparison does not certify
the remaining outputs.  Natural-task F1 is separate evidence: it measures
user-visible utility but cannot establish internal equality.  The principal
experiment's LRU64-versus-zero-cache comparison tests cache-policy equality
within its recorded executable.  A separate zero-cache source-path canary compares
an eight-window direct source against its synchronous-source control on a
256-token prompt and eight-output horizon.  It compares all eight complete
logit rows, tokens, response bytes, 384 ordered route events, consumer and
destination identities, and clean closure.  Its one prefill event at each of
48 layers selects all 512 experts, covering all 24,576 managed layer--expert
objects.  This is an eight-output source-path result, not part of the 64-output
cache-policy horizon.  Equality to an unmodified upstream runtime and direct
recurrent/hybrid-state bytes are unavailable and not claimed.

This distinction matters for sparse and hybrid models.  Router decisions
amplify numeric differences into future I/O; recurrent state differences can
alter later speculative acceptance.  The implementation retains native quantized bytes and
the existing graph so that a source-path comparison is not also a model-
conversion experiment.

\subsection{Lifetime Is Part of Correctness}

A reusable resource version must not be published before its movement
completes, must denote the requested object and version, and must not be reused
before declared consumers finish.  Reset and shutdown must drain or invalidate
outstanding work, and an invalid completion must not make data visible.  These
properties do not prescribe a bookkeeping mechanism.  The implementation realizes them
with request epochs, slot generations, readiness events, tickets, and consumer
counts.  The reported evidence covers tested implementation behaviors, not a
formal proof of race freedom.

\section{Design}
\label{sec:design}

\begin{figure*}[t]
  \centering
  \begin{tikzpicture}[
    font=\small,
    box/.style={draw,rounded corners,minimum height=1.05cm,text width=2.65cm,align=center,fill=blue!6},
    ctl/.style={draw,rounded corners,minimum height=0.9cm,text width=2.65cm,align=center,fill=orange!8},
    cert/.style={draw,dashed,rounded corners,inner sep=8pt},
    flow/.style={-{Latex[length=2mm]},thick},
    cflow/.style={-{Latex[length=2mm]},dashed}
  ]
    \node[box] (store) {Canonical tensor extents\\native qtypes};
    \node[box,right=1.05cm of store] (windows) {Fixed host windows\\bounded direct I/O};
    \node[box,right=1.05cm of windows] (slots) {Reusable GPU slots\\stable addresses};
    \node[box,right=1.05cm of slots] (graph) {Native model graph\\kernels, router, sampler};
    \draw[flow] (store) -- node[above,font=\scriptsize,fill=white,inner sep=1pt]{aligned read} (windows);
    \draw[flow] (windows) -- node[above,font=\scriptsize,fill=white,inner sep=1pt]{H2D} (slots);
    \draw[flow] (slots) -- node[above,font=\scriptsize,fill=white,inner sep=1pt]{consume} (graph);

    \node[ctl,below=1.15cm of windows] (plan) {Shared \Plan\\objects, extents, generations};
    \node[ctl,below=1.15cm of slots] (life) {Ticketed lifecycle\\epochs and consumers};
    \draw[cflow] (plan) -- (store);
    \draw[cflow] (plan) -- (windows);
    \draw[cflow] (plan) -- (slots);
    \draw[cflow] (life) -- (windows);
    \draw[cflow] (life) -- (slots);
    \draw[cflow] (life) -- (graph);

    \node[cert,fit=(store)(windows)(slots)(graph)(plan)(life),label={[font=\footnotesize]below:
      complete cgroup host accounting $+$ whole-board GPU observation $+$ identity ledger}] {};
  \end{tikzpicture}
  \caption{The architecture separates canonical ownership, bounded physical residency,
  stable execution addresses, and scheduling.  The model graph consumes native
  tensors at stable slots; the \Plan and ticket protocol determine which bytes
  those addresses denote.}
  \label{fig:architecture}
\end{figure*}
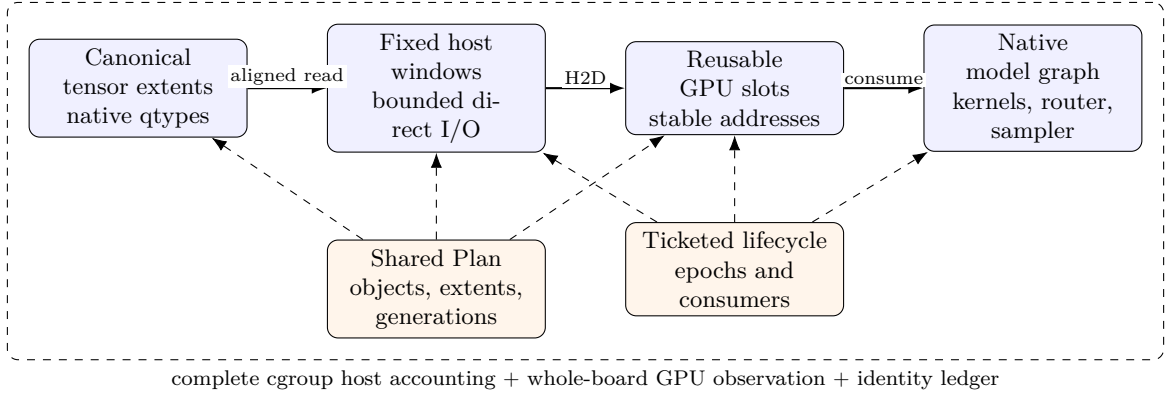

Figure~\ref{fig:architecture} separates the data and control paths.  Canonical
tensor extents flow through fixed host windows into reusable GPU slots.  The
control path resolves a model-layer request to an object bundle, initiates
bounded reads and copies, and publishes stable execution addresses after the
tested identity and readiness checks.  External observers account for the
resource envelope independently of the model process.

\subsection{A Shared Cross-Adapter Plan Core}

The \Plan comprises an immutable manifest and mutable execution state.  The
manifest records object identifiers, logical sizes, source extents, alignment
and padding, compatible windows and slots, and stable device addresses.  Object
identifiers express model semantics rather than file layout.  An adapter can
therefore combine several extents into one logical object without exposing the
container layout to the graph.

Mutable state associates each window and slot with an object, generation,
request epoch, readiness event, and live-consumer count.  Replacement policy is
separate.  Demand loading, LRU, prediction, and pinning select different future
objects but use the same lifecycle interface.  Transfer between Qwen and Gemma
changed object mapping and adapter code; it did not establish a general
abstraction theorem.

\begin{table}[t]
  \centering
  \scriptsize
  \caption{Shared-core and adapter responsibilities.  Cross-adapter reuse does
  not imply unrestricted model neutrality.}
  \label{tab:interface}
  \begin{tabularx}{\columnwidth}{@{}p{0.43\columnwidth}X@{}}
    \toprule
    Shared \Plan core & Model/runtime adapter \\
    \midrule
    Extents and slot compatibility & Container object discovery \\
    Versioned resource state & Bundle and layer-demand semantics \\
    Readiness and publication & Graph and kernel integration \\
    Consumer and recycle lifetime & Router, KV, and recurrent state \\
    Reset/shutdown invalidation & Model-specific oracle surfaces \\
    Telemetry and validators & Native layout interpretation \\
    \bottomrule
  \end{tabularx}
\end{table}

\subsection{Bounded Source Service}

The source service opens canonical files for direct I/O and preallocates a
fixed set of aligned host windows.  It computes covering reads, submits no more
than the configured in-flight limit, and records logical bytes, physical bytes,
padding, completion status, and latency.  Short reads, negative completions,
extent overflow, and object-size disagreement are rejection conditions in the
tested implementation.  Padding is excluded from logical tensor bytes.

Each completion carries the object, extent, window, request epoch, and
generation present at submission.  A completion with a mismatch does not
publish host readiness in covered unit tests.  After all extents are assembled,
the window becomes eligible for H2D.  Fixed windows give the explicit bound

\begin{equation}
M_{\mathrm{source}} \leq N_w W + M_{\mathrm{queue}} + M_{\mathrm{meta}},
\end{equation}

where $N_w$ is the number of windows and $W$ their aligned size.  Direct I/O
avoids a second copy of the streamed payload accumulating in the page cache;
complete cgroup accounting remains necessary for every other host allocation.

\subsection{Stable Slots and Bundle Publication}

Model runtimes often capture tensor addresses in graph nodes or device
descriptors.  The implementation reserves a bounded slot pool whose addresses remain
stable while logical identity and generation change.  Before a layer executes,
an adapter requests its object bundle.  In the tested implementation, all
components are checked and reserved before one publication state transition
makes the bundle visible.  Unit negative controls and real-adapter deterministic
cells cover short and failed reads, partial bundles, out-of-order completion,
stale generations, wrong objects, held consumers, reset, and shutdown.  This is
not described as transactional: untested schedules and races remain open.

\subsection{Separate Window and GPU-Slot Lifecycles}

Host windows and GPU slots are different resources.  Their state sequences are
therefore specified separately:

\begin{align*}
\text{window: }&\textsc{Free}\rightarrow\textsc{Reading}
 \rightarrow\textsc{HostReady}\\
&\hspace{3.6em}\rightarrow\textsc{Copying}\rightarrow\textsc{Free},\\
\text{GPU slot: }&\textsc{Vacant}\rightarrow\textsc{Loading}
 \rightarrow\textsc{DeviceReady}\\
&\hspace{3.6em}\rightarrow\textsc{InUse}\rightarrow\textsc{Recyclable}
 \rightarrow\textsc{Vacant}.
\end{align*}

A host window can return to \textsc{Free} after its H2D dependency completes;
a GPU slot cannot become \textsc{Recyclable} until the copy completes and its
generation has zero live consumers.  Tickets carry object, generation, epoch,
and resource identity across the two state machines.  Reset advances the epoch
after the tested outstanding-work checks; shutdown also checks submitted source
operations.  Covered invalid transitions become explicit errors.  The tested
implementation fails closed for the named deterministic cells in
Section~\ref{sec:faults}; no claim extends to all integrated faults.

\subsection{Temporal and Spatial Amortization}

The same service boundary supports two distinct economies.  \emph{Spatial
amortization} retains recently used objects so future traversals avoid storage
and H2D traffic.  \emph{Temporal amortization} commits several output tokens
per target traversal using a proposer, while exact state rebasing preserves the
target trajectory.  Cache capacity and multi-token acceptance are reported as
different mechanisms even when they share storage and state infrastructure.

\section{Implementation}
\label{sec:implementation}

The prototype contains a C++ \Plan core, a bounded source service, coordination
for host windows and GPU slots, claim-level telemetry, and model-specific
adapters.  Qwen3-Coder AWQ campaigns use a pinned vLLM-derived runtime; the
Qwen3.6, Gemma, and Qwen3-Next campaigns use pinned \texttt{llama.cpp}-derived
lines~\cite{llamacpp}.  Each campaign has a separate source and runtime identity.
The paper does not imply that results obtained in one runtime were reproduced
in the other.

\para{Object discovery}
Adapters inventory exact native tensor extents at load time.  Qwen AWQ, Qwen3.6
Q4\_K\_M, Gemma QAT Q4\_0, and Qwen3-Next Q4\_K\_M use different naming and
bundle rules.  Logical and aligned physical bytes are recorded separately, so
padding does not inflate the capacity numerator.

\para{Scheduling and copies}
Layer entry requests a bundle.  Misses are submitted to the source service,
completed windows are copied into reserved slots, and a device event publishes
readiness.  Layer exit releases the consumer ticket.  Host-window count and
GPU-slot capacity are independent; the Qwen parity diagnostic uses 144 windows,
eight slots, and 3,384 scheduled objects per request.

\para{Temporal state}
The Qwen3.6 path couples one target with its linear MTP block and seven physical
branch states.  Winner commitment and six loser rebases preserve shared hybrid
state.  The oracle compares every recorded target logit row, all committed
tokens, shared-prefix hashes, and state equality after each commit.

\para{Telemetry}
The service reports submissions and completions, in-flight depth, logical and
physical source bytes, padding, H2D bytes, cache events, generations, and
lifecycle rejections.  The outer harness records cgroup membership and events,
whole-board memory and compute-process ownership, output digests, and phase
times.  A successful process exit alone is never a claim predicate.

\para{Artifact}
An audit and replay artifact accompanies this report.  It preserves exact
claim-bearing experimental source snapshots, reconstructs each patched runtime
from a public upstream commit and patch series, and keeps portability helpers
separate from the recorded experiment files.  Audit verifies C1--C12 record
identities and enumerated assertions offline; replay recomputes arithmetic and
categorical summaries from packaged compact operands.  Neither operation
rederives every certificate from raw execution output.  The principal C1--C3
source and execution contracts specify inputs for new executions, each with its
own identity.  Model weights and bulk binary outputs are not redistributed;
pinned acquisition metadata, byte lengths, expected hashes, compact
observations, and validators are supplied.  The stable interface is
\texttt{./artifact verify --claim C\#} and
\texttt{./artifact replay --claim C\#}.

\begin{table*}[t]
  \centering
  \scriptsize
  \caption{Principal results by evidence identity.  ``Host-hard/GPU-audited''
  does not imply a hard GPU partition; the physical board is a separate
  capacity ceiling.  Performance values are not pooled across rows.}
  \label{tab:summary}
  \begin{tabularx}{\textwidth}{@{}p{2.45cm}p{2.55cm}p{3.55cm}X@{}}
    \toprule
    Campaign / system & Resource authority & Exactness scope & Principal result \\
    \midrule
    \textbf{Demanded weights \claimref{C1}}\\Qwen3-Next Q4\_K\_M; RTX 3090 & Host hard 11\GiB; GPU audited 23\GiB on physical 24\GiB board & $E_{\mathrm{64-out}}$: all 64 complete logit rows, tokens, response bytes, routes, and identities exact versus zero-cache & $T_{\mathrm{repr}}=45.08$\GiB; $U_{\mathrm{sem}}^{\mathrm{LB}}=43.59375$\GiB exceeds the declared 34\GiB envelope by 9.59375\GiB; 32K+64 completes with zero resource events. \\
    \textbf{Matched source \claimref{C2}}\\same Qwen3-Next/device & Host hard 11\GiB; GPU audited 23\GiB & Six exact, resource-clean D/F pairs; P exact but at the host limit with 33,481 limit events & Complete asynchronous-component F/D wall ratio 0.3229; all six ratios lie in [0.3211, 0.3283]; physical bytes/output ratio 1.0; churn predicate fails. \\
    \textbf{Integrated faults \claimref{C3}}\\later Qwen3-Next and Gemma 4 builds; RTX 3090 & Qwen hard host 11\GiB; Gemma hard host 8\GiB; zero swap/events & All fourteen prespecified cells satisfy their validators; fatal cells publish nothing invalid; recoverable cells control-exact & Fail-closed behavior for named transitions across two real adapters in the tested binaries; not attribution to the earlier C1 executable or race freedom. \\
    \textbf{Bounded source \claimref{C8}}\\same Gemma/device & Hard host 8\GiB; 12\GiB board observed & Full logits, tokens, routes, and cache state exact in one pair & Host peak 6.19 to 2.78\GiB; 14,396 balanced direct-I/O submissions and completions. \\
    \bottomrule
  \end{tabularx}
\end{table*}

\section{Evaluation}
\label{sec:evaluation}

The evaluation asks five questions:

\begin{enumerate}[leftmargin=*]
  \item Can native execution remain full-output exact when the workload's
  semantic-demand lower bound exceeds both the declared full-residency envelope
  and the host-hard/device-physical envelope?
  \item Relative to the tested buffered path, does direct ownership change the
  prespecified positive-margin/no-limit-event admission outcome?
  \item Against a blocking one-window direct reader, how does the complete
  eight-window asynchronous component change wall time at identical physical
  source bytes?
  \item Do named invalid source and ownership transitions fail closed through
  two real adapter/GPU paths?
  \item Does the source boundary transfer to a second adapter, and does
  resource admissibility imply product superiority?
\end{enumerate}

Table~\ref{tab:summary} reports the evidence populations separately.  Each row
has its own model, runtime, workload, oracle, resource contract, cache protocol,
and repetition rule.

\subsection{Methodology and Evidence Cutoff}

The evidence cutoff includes the Qwen3-Next matched-source and
deterministic-fault populations, the Gemma representative fault transfer, the
Qwen3-Next full-output oracle confirmation, and its demand-accounting
derivation, all dated no later than 21 August 2026.  Later results are outside
this report and cannot support its claims.

\para{Principal evidence records}
The principal results are detailed in
Sections~\ref{sec:aggregate}--\ref{sec:faults}.  C1 is an earlier-executable
\textsc{W}/$\textsc{E}[E_{\mathrm{64-out}},64]$ certificate with clean
no-fault lifecycle closure.  C2 is a distinct matched-source admission and
complete-component population.  C3 is a later-build $\textsc{L}[F]$
transition validation across two adapters.  These records form one evidence
package but are not pooled into one executable certificate.  Temporal,
Qwen-specific, shared-\Plan, and Gemma cache evidence appears in
Appendix~\ref{sec:additional} at its own identities.

\begin{table*}[t]
  \centering
  \scriptsize
  \caption{Executable identity boundary for the principal results.  The \Plan
  state-machine source is byte-identical across these rows (SHA-256 prefix
  \texttt{131c5d0036e0}); adapter and bounded-source identities are not.  A
  passing row therefore does not transfer untested behavior to another binary.}
  \label{tab:identities}
  \begin{tabularx}{\textwidth}{@{}p{1.7cm}p{2.1cm}p{2.25cm}p{3.0cm}X@{}}
    \toprule
    Record & Workload / role & Executable SHA-256 & Adapter / source SHA-256 & Direct attribution \\
    \midrule
    C1 & Qwen 32K+64 principal experiment & \texttt{4c68fc99e72c} & \texttt{ca637a7c84e3} / \texttt{7a2d224375a5} & Full-output exactness, resources, and clean no-fault closure. \\
    C2 & Qwen 8K+128 matched P/D/F & \texttt{7bd52530b6ad} & \texttt{b126d4b58827} / \texttt{a5f1453eb07e} & Source admission and six-pair complete-component result. \\
    C3-Q & Qwen 512+8 named transitions & \texttt{5083ccd5b1fa} & \texttt{029ed3af631d} / \texttt{1c74a7cfba92} & Ten later-build control/fault cells. \\
    C3-G & Gemma named-transition transfer & \texttt{a01ea9314823} & \texttt{88570b5db5a7} / \texttt{1c74a7cfba92} & Four later-build control/fault cells. \\
    \bottomrule
  \end{tabularx}
\end{table*}

\para{Machines}
Local RTX 3090 campaigns ran on a workstation with an AMD Ryzen 7 5800X
(8 cores/16 threads), a 2\,TB WD\_BLACK SN850X NVMe device, a physical 24\GiB
RTX 3090, and NVIDIA driver 580.159.03.  RTX 3080 Ti campaigns ran on one
physical 12\GiB board with driver 580.178.04; the measurement record identifies
the source filesystem as \texttt{/dev/vda1} and the backing leaf device as
\texttt{vda}.  The remote measurement record does not identify the physical CPU
model or the provider's underlying storage medium.  Accordingly, the report
makes no CPU-normalized or physical-storage-normalized performance claim for
the RTX 3080 Ti rows.  A public replay must capture those two identities.

\para{Resource observation}
Host measurements use cgroup \texttt{memory.max}, \texttt{memory.peak},
swap, and event counters over the complete process hierarchy.  Memory peak is a
kernel high-water mark rather than a sampled estimate.  Whole-board GPU and
foreign-process observations use 100\,ms sampling for Qwen AWQ campaigns and
50\,ms sampling for the Gemma and Qwen3-Next campaigns.  The declared GPU
budget is compared with the maximum whole-board observation; no experiment
uses a hard per-process GPU partition.

\para{Cache and storage state}
Qwen natural utility compares the unchanged protected-LRU64 plus two-bank
next-layer-prefetch mechanism with its established control; it is not a \Plan
campaign.  Each task process contains primary and immediate-warm cycles.  The
separate \Plan parity process also contains primary and warm requests with reset
between them.  Gemma cache pairs start with a cold prefill and compare zero
reuse with independent per-layer LRU32, using prefill-tail seeding and
alternating fixed order across five prompts.  The bounded-source pair holds the
Gemma capacity-32 path fixed and substitutes eight aligned O\_DIRECT windows.
Qwen3-Next uses per-layer LRU64; its long prefill warms the subsequent short
decode.  The later matched-source experiment holds LRU64 fixed across P, D, and F;
fault cells disable cache and use eight fixed O\_DIRECT windows.  Gemma fault
cells use the same zero-cache, eight-window source boundary.  MTP uses two
256\MiB service slots and greedy target/proposer sampling.

\para{Repetitions}
The Qwen natural population has eight control and eight candidate processes;
the two request cycles within a process are not independent repetitions.  The
\Plan comparison has 16 valid cells and was not independently repeated.  The
\Plan parity diagnostic is one process.  Gemma cache has one observation per
arm and prompt; bounded source is one matched pair.  The Qwen3-Next 32K+64
execution and the Qwen3.6 complete MTP request each have one claim-bearing
observation after a smaller validation run.  The full-64 confirmation uses one
prespecified zero-cache comparator, zero statistical repetitions, and zero
reruns.  The matched Qwen3-Next source
campaign has six preordered counterbalanced D/F pairs and a fixed paired-log
$t$ interval; P descendants stop after its first valid positive-margin/
no-limit-event admission failure.  That
interval quantifies repetition at one fixed model, prompt, horizon, runtime,
device, and storage identity.  It is not an interval over prompts, models,
hardware, or deployments.  All fourteen prespecified deterministic fault
cells are one outcome-blind execution per prespecified mode, not statistical
reliability repetitions.  Other timing values without
an independent repetition are descriptive and have no variance estimate.

\long\def\additionalexperiments{%
\section{Additional Experiments and Transfer Evidence}
\label{sec:additional}

\para{Scope}
The next five subsections report earlier and cross-model experiments related to
the principal mechanisms.  Each retains its own oracle, workload, runtime,
resource contract, and repetition design.

\begin{table*}[t]
  \centering
  \scriptsize
  \caption{Additional experiment populations.  Values remain scoped to their
  own identities and are not pooled with the principal Qwen3-Next experiment.}
  \label{tab:additional}
  \begin{tabularx}{\textwidth}{@{}p{2.6cm}p{3.0cm}p{4.0cm}X@{}}
    \toprule
    Campaign & Resource authority & Correctness or utility scope & Result \\
    \midrule
    Qwen3.6 temporal MTP \claimref{C4} & Host 7.5\GiB; GPU 11.5\GiB & 723
      full-logit rows, 512 tokens, and exposed hybrid state exact at one
      8,192+512 identity & One prefill-produced output plus 511 decode outputs
      in 104 target traversals; 4.913 traversed outputs/traversal; 12.016
      end-to-end tokens/s. \\
    Qwen-specific mechanism \claimref{C5} & Host hard 6\GiB; 12\GiB board
      audited & Eight LongBench-derived tasks; semantic F1; 15/16 request
      outputs token-identical & Equal aggregate F1 0.590909; Qwen-specific
      prefetch cycle 331.884 to 235.917 s ($-28.916\%$). \\
    Shared-\Plan parity \claimref{C12} & Host hard 6\GiB; board audited & One
      8,192+32 primary/warm process; token digests and lifecycle & Exact parity
      and balanced 144-window/eight-slot teardown; no performance claim. \\
    Shared-\Plan comparison \claimref{C6} & Host hard 6\GiB; board audited &
      Sixteen valid cells; prespecified F1 criterion fails & Request+reset
      335.400 to 250.785 s ($-25.228\%$); aggregate F1 delta
      $-0.0015625$ remains a failure. \\
    Gemma cache transfer \claimref{C7} & Host hard 8\GiB; 12\GiB board audited
      & Full logits and tokens exact in five same-scheduler pairs & H2D bytes
      $-65.021\%$; process physical reads $-61.474\%$; decode wall
      $-77.517\%$. \\
    \bottomrule
  \end{tabularx}
\end{table*}

\subsection{Temporal Amortization with Exact Hybrid-State Rebasing}
\label{sec:mtp}

The temporal experiment uses a canonical Qwen3.6-27B Q4\_K\_M target and its
linear MTP block in one composite GGUF~\cite{qwen3report}.  The natural
retrieval request has 8,192 prompt tokens and 512 generated outputs.  Greedy
target and proposer execution uses seven lanes, a maximum draft length of six,
Q4\_0 KV, and two 256\MiB \Plan service slots.  The candidate request excludes
oracle time from its wall clock.

The request completes in 42.611 s, including 7.608 s to first token, for 12.016
end-to-end output tokens/s and 14.599 generation-phase tokens/s.  It requires
104 target traversals, accepts 407 draft tokens, and fully accepts 51 draft
groups.  After the prefill-produced first output, the 511 decode outputs give
4.913 committed outputs per target traversal\claimref{C4}.  The
state ledger records 104 winner commits, 624 loser rebases, and 728 state
stages.  The weight service balances 4,032 mappings and unmaps, 4,030 recycles,
26,712 consumer completions, and one shutdown.

The same-build oracle checks 723 complete logit rows, every output token, state
equality after every commit, winner/loser rebasing, and shared-prefix
immutability.  All pass bitwise at the scoped identity.  Complete-process-tree
host peak is
8,050,614,272 bytes under 8,053,063,680; GPU peak is 12,335,448,064 bytes under
the declared 12,348,030,976-byte ceiling.  Both margins are positive but narrow
(2.45\,MB host and 12.58\,MB GPU).  This is one developed prompt and one
microbatch-dependent floating-point trajectory.  It establishes temporal
amortization at that identity, not a cross-prompt MTP law or a comparison with
an external 8.803 tokens/s point.

\subsection{Qwen-Specific Natural Utility Before \Plan Extraction}
\label{sec:qwennatural}

The first Qwen AWQ population evaluates the unchanged protected-LRU64 plus
two-bank next-layer-prefetch mechanism on eight support-preserving,
output-blind LongBench-derived QA records~\cite{longbench}.  It uses 8,192 input
tokens, at most 128 outputs, and a hard 6\GiB complete-process-tree host
boundary.  This is a
Qwen-specific mechanism result, not shared-\Plan evidence.

Across eight control and eight candidate processes, both arms obtain aggregate
F1 0.590909 and all eight task-level F1 values agree.  Fifteen of sixteen
request outputs are token-identical; the differing NarrativeQA warm output
scores zero in both arms.  Request-plus-reset cycles fall from 331.884 to
235.917 s, a 28.916\% reduction, with all eight task directions
favorable\claimref{C5}.  The
mechanism's output work differs by four tokens across the population and is
part of delivered utility.  The supported claim is equal aggregate F1 and
favorable observed cycle time on this population, not exact-trace causality or
a \Plan speedup.

\subsection{Shared-\Plan Parity Diagnostic}
\label{sec:qwenparity}

Extraction into the shared \Plan is tested in a separate deterministic
8,192+32 process.  It schedules 3,384 objects per request through 144 host
windows and eight GPU slots, then performs reset, an immediate-warm request,
reset, and shutdown.  Primary and warm token digests equal the inherited
same-build oracle; requests, resets, windows, slots, consumers, and tickets are
balanced at teardown.  Host peak is 5,013,585,920 bytes under 6\GiB.  Whole-
board peak is 12,294,094,848 bytes on the 12\GiB RTX 3080 Ti, leaving the
declared 590,807,040-byte margin.

This one process is implementation-parity evidence.  It is not a natural-task
population, multi-cell utility comparison, or a performance
result\claimref{C12}.

\subsection{Qwen \Plan Comparison and Prespecified F1 Miss}
\label{sec:qwencomparison}

This comparison is a third Qwen identity: 16 complete valid cells on the
same model family and physical RTX 3080 Ti.  The conservative nonresident
qweight population is 7,247,757,312 bytes, 805,306,368 bytes larger than the
hard 6\GiB host cap.  Complete resource ownership, scoped \Plan
prefill/migration ownership, reset, lifecycle, and teardown pass.

The strongest control satisfying the exactness contract requires 335.400 s of aggregate
request-plus-reset time; the \Plan candidate requires 250.785 s, a 25.228\%
reduction.  This value is distinct from the 28.916\% Qwen-specific natural
mechanism result\claimref{C6}.  The prespecified official-LongBench zero-margin F1 criterion
is not met: one task changes the eight-direction aggregate by $-0.0015625$.  The
candidate answer contains the exact reference ``Eindhoven,'' equals a valid
control answer, and lies within larger cold/warm control variation.  No
deterministic \Plan-specific error was identified, but the prespecified utility
criterion failed and the cause remains unresolved.  The numerical miss is not
reclassified as a pass.

\subsection{Gemma Transfer and Cache Mechanism}
\label{sec:gemma}

Gemma 4's official QAT Q4\_0 representation supplies the second model-family
path~\cite{gemma4report}.  Five prompt pairs compare zero reuse with independent
per-layer LRU32 under a hard 8\GiB cgroup and whole-board observation of the
12\GiB RTX 3080 Ti.  Full logits and tokens match within all five same-scheduler,
same-representation pairs.  This does not assert identity to ordinary unsplit
\texttt{llama.cpp}.

Across 501 generated tokens, H2D traffic falls from 462,068,305,440 to
161,626,889,956 bytes, a 65.021\% reduction.  Process block reads fall 61.474\%;
because page cache, readahead, and amplification mediate this counter, it is not
a universal disk-bandwidth claim.  Aggregate decode time falls from 1,199.513
to 269.682 s, or 77.517\%, with all five directions
favorable\claimref{C7}.  One observation
per cell and alternating order reduce neither storage-state nor timing
uncertainty to a confidence interval.

Worst complete-process-tree host peak is 6,730,268,672 bytes, leaving
1,859,665,920 bytes.
Worst whole-board peak is 6,065,553,408 bytes.  Swap and resource events are
zero.  A same-build partial-offload comparator completes generation but reaches
the 8\GiB host boundary, with 10,325 \texttt{memory.max} events under mmap and
7,922 without mmap; it fails the same positive-headroom/no-limit-event admission
predicate under this exact host contract.  Only
three prompts complete their task and two reach natural EOS, so the population
supports mechanism transfer and traffic reduction rather than general semantic
quality.

}

\long\def\boundedtransfer{%
\subsection{Bounded Direct I/O Establishes Host Nonresidency}
\label{sec:boundedsource}

A matched Gemma pair replaces the current buffered source with the eight-window
bounded O\_DIRECT executor while retaining the model, graph, cache capacity,
and scheduler.  The managed expert inventory is
$S=12{,}846{,}382{,}080$ bytes.  With
$H_{\mathrm{cap}}=8{,}589{,}934{,}592$ and declared candidate reserve
$R_{\mathrm{policy}}=2{,}935{,}803{,}872$, target capacity while preserving the
reserve is $H_{\mathrm{cap}}-R_{\mathrm{policy}}=5{,}654{,}130{,}720$.
Equation~\ref{eq:hostnonresident} therefore holds by 7,192,251,360 bytes; this
is a policy-relative nonresidency statement, not a lower bound on unavoidable
runtime state.

The bounded source reduces complete-process-tree host peak from 6,641,348,608 to
2,980,458,496 bytes (55.123\%).  It issues and completes exactly 14,396 reads,
with peak depth eight, zero cancellation or failure, and all eight windows free
at shutdown.  Cumulative logical traffic is 48,160,551,152 bytes; aligned physical reads
are 48,336,400,384 bytes, including 175,849,232 padding bytes.  Full logits,
tokens, routes, and cache state match the control.  Total wall time falls from
133.645 to 62.021 s in this single pair; the physical bound and balanced ledger,
not this unrepeated timing difference, are the principal
result\claimref{C8}.

}

\subsection{Output-Exact Execution Beyond the Full-Residency Envelope}
\label{sec:aggregate}

\begin{table}[t]
  \centering
  \footnotesize
  \renewcommand{\arraystretch}{1.08}
  \caption{Principal C1 certificate instance.  C3 is a separate later-build
  transition record and is not a conjunct of C1.}
  \label{tab:c1certificate}
  \begin{tabularx}{\columnwidth}{@{}p{0.10\columnwidth}X@{}}
    \toprule
    Field & C1 instantiation \\
    \midrule
    $I$ & Qwen3-Next Q4\_K\_M, 32K+64, executable
      \texttt{4c68fc99e72c}; 45.08\GiB representation and
      43.59375\GiB semantic-demand lower bound. \\
    $B,A$ & Host 11\GiB hard cgroup-charge boundary; GPU 23\GiB
      whole-board audit on a physical 24\GiB board. \\
    $E$ & One same-build zero-cache comparator; exact
      $E_{\mathrm{64-out}}$ over 64 outputs. \\
    $L$ & Ordinary publication, movement, recycle, consumer release,
      reset/shutdown, and clean no-fault closure.  Fault cells excluded. \\
    \bottomrule
  \end{tabularx}
\end{table}

The principal experiment uses Qwen3-Next-80B-A3B-Instruct in native Q4\_K\_M
\cite{qwen3next,qwen3report}.  Its complete canonical tensor representation is
$T_{\mathrm{repr}}=48{,}405{,}005{,}312$ bytes (45.08\GiB).  The process tree has
a hard $H_{\mathrm{cap}}=11$\GiB cgroup limit.  The GPU contract declares and
audits $G_{\mathrm{decl}}=23$\GiB of whole-board use on a physical
$G_{\mathrm{phys}}=24$\GiB RTX 3090.  Hence

\begin{align*}
T_{\mathrm{repr}}-(H_{\mathrm{cap}}+G_{\mathrm{decl}})
  &=11{,}897{,}783{,}296\ \mathrm{bytes}.
\end{align*}

The representation therefore exceeds the declared 34\GiB full-residency
envelope by 11.08\GiB.  This alone would not show that the workload touches the
oversized portion.  The recorded prefill trace closes that alternative: the
tested adapter copies each completed stock \texttt{ffn\_moe\_topk} tensor
before any \Plan scheduling or cache action.  All eight chunks contain 48 layer
events, and every event's authoritative router output contains all 512 experts.
All 24,576 layer--expert identities therefore have at least one routed token.
For each selected expert, the native dense expert path invokes the complete
gate, up, and down projections; those three canonical extents therefore enter
the semantic-demand union.

The audit joins those identities to each native gate, up, and down expert
extent.  Its 73,728 component intervals have zero duplicates and zero overlaps
under the key (model digest, source offset, logical length).  The sorted extent-
union digest is \texttt{bad5b23f5ac0...}; its deduplicated size is
$U_{\mathrm{sem}}^{\mathrm{LB}}=46{,}808{,}432{,}640$ bytes
(43.59375\GiB).  This conservative semantic-demand lower bound excludes the
remaining 1.4869\GiB of canonical tensors rather than inferring their demand.
Thus

\begin{align*}
U_{\mathrm{sem}}^{\mathrm{LB}}-(H_{\mathrm{cap}}+G_{\mathrm{decl}})
  &=10{,}301{,}210{,}624\ \mathrm{bytes},\\
U_{\mathrm{sem}}^{\mathrm{LB}}-(H_{\mathrm{cap}}+G_{\mathrm{phys}})
  &=9{,}227{,}468{,}800\ \mathrm{bytes}.
\end{align*}

The semantic-demand lower bound exceeds the declared envelope by 9.59375\GiB
and the host-hard/device-physical envelope by 8.59375\GiB\claimref{C1}.
``Demanded'' means selected for at least one token, not simultaneously live.
Scheduler requests and candidate cumulative source traffic---392,942,788,608
logical bytes and 395,477,815,296 aligned physical bytes across rereads---are
separate quantities and are not used as the semantic-demand numerator.  The
Linux-visible whole-machine observation is reported in
Section~\ref{sec:contract}; it is not a claim about all installed host DRAM.

One prespecified zero-cache 32K+64 oracle uses the same tested
executable, model, prompt, graph, greedy sampler, source path, hardware, and
resource contract as the fixed LRU64 experiment.  Every byte of all 64
complete 151,936-float logit rows is equal (38,895,616 bytes per arm), as are
all 64 token IDs, response bytes, all 3,408 recorded route events, and the
seven-field native-consumer and two-entry model-memory destination records.
Their compact, sorted-key JSON SHA-256 digests are respectively
\texttt{f5ae6bc54add...} and \texttt{e23ab3127932...} in both arms, and the
validator also compares the records directly.  The experimental harness does not capture
recurrent/hybrid-state bytes, so output equality is not promoted to a direct
state-byte claim.  A separate complete source-path equivalence test is also
bitwise exact at the 256+8 scope defined in
Section~\ref{sec:contract}, but it is not pooled with the cache-policy comparison.

The principal cell processes a 32,768-token mechanics prompt and generates 64
tokens, with 63 measured decode transitions, per-layer LRU64, and a warm decode
following the long prefill.  Prefill takes 86.355 s and decode 6.454 s, or 9.762
transitions/s.  All 64 tokens execute through the native graph, source and
destination counts balance, and reset/teardown complete.  Complete-process-tree
cgroup-charge peak is
1,496,981,504 bytes; whole-board peak is 13,218,807,808 bytes, leaving
11,477,254,144 bytes below the declared GPU budget.  Swap, OOM,
\texttt{memory.max}, and foreign-compute events are zero.

A 512+64 mechanism diagnostic decodes at 5.037 tokens/s without reuse and
11.636 tokens/s with LRU64.  It explains why retention matters at this identity
but is not a policy study.  The principal experiment remains one model, one device, one
short warm-cache observation, and a mechanics prompt composed of four repeated
copies of a held-out code fixture.  It establishes resource admissibility under
the prespecified hard-margin contract and full-output exactness at this 64-output
identity, not semantic quality, longer-horizon stability, direct recurrent-state
equality, or hardware generality.

\subsection{Matched Source Ownership and Asynchronous Execution}
\label{sec:matchedsource}

The matched ownership and asynchronous-component experiment holds the
Qwen3-Next model bytes, graph, sampler,
per-layer LRU64, 8,192-token natural prompt, 128 greedy outputs, object order,
11\GiB cgroup, and 23\GiB GPU contract fixed.  P serves extents through the
kernel page cache; D uses blocking O\_DIRECT and one fixed window; F uses eight
registered windows, bounded asynchronous O\_DIRECT, and the complete \Plan
ticket, generation, consumer, reset, and shutdown path.  A separate same-build
synchronous-source oracle supplies all 128 full-vocabulary logit rows, tokens,
pieces, routes, source identities, and final cache state.

The first valid P cell is oracle-exact but reaches the 11\GiB ceiling and
advances \texttt{memory.max} 33,481 times.  It therefore fails the
prespecified positive-headroom/no-limit-event admission predicate; this does
not mean that it failed to complete.  Its descendants stop as specified.  All
six preordered D/F pairs are oracle-exact, pass that predicate, and are
lifecycle-balanced.  Mean complete wall is 101.739 s for D and 32.852 s for F.
The paired geometric F/D wall ratio is 0.322913, with a 95\% paired-log $t$
interval of [0.320090, 0.325762].  The F/D exposed-source-wait ratio is 0.534673,
decode-rate ratio is 1.541885, and physical source bytes per output are
identical (ratio 1.0).  Every F cell issues and completes 55,520 source
operations at peak depth eight, performs zero dynamic direct allocation, and
returns all windows free.

\begin{table}[t]
  \centering
  \footnotesize
  \renewcommand{\arraystretch}{1.08}
  \caption{All six matched complete-wall pairs and their actual D/F execution
  order.  Ratios repeat one fixed workload/system identity; they do not sample
  prompts or hardware.}
  \label{tab:matchedpairs}
  \begin{tabular}{@{}crrrr@{}}
    \toprule
    Block & Order & D (s) & F (s) & F/D \\
    \midrule
    1 & $\mathrm{D}\!\rightarrow\!\mathrm{F}$ & 102.358 & 32.867 & 0.321095 \\
    2 & $\mathrm{D}\!\rightarrow\!\mathrm{F}$ & 102.228 & 32.820 & 0.321052 \\
    3 & $\mathrm{F}\!\rightarrow\!\mathrm{D}$ & 100.032 & 32.837 & 0.328269 \\
    4 & $\mathrm{F}\!\rightarrow\!\mathrm{D}$ & 101.996 & 32.889 & 0.322449 \\
    5 & $\mathrm{D}\!\rightarrow\!\mathrm{F}$ & 102.202 & 32.887 & 0.321784 \\
    6 & $\mathrm{F}\!\rightarrow\!\mathrm{D}$ & 101.616 & 32.811 & 0.322888 \\
    \bottomrule
  \end{tabular}
\end{table}

The narrow interval measures repeatability across these six executions at the
fixed identity.  It is not a confidence interval for generalization across
prompts, models, storage devices, or hardware.

The prespecified workload-churn premise does not hold.  Decode produces 6,471
aggregate generations rather than 9,216, and the weakest layer produces 61
rather than 128.  Although 3,063 of 3,072 slots recycle, the conjunctive churn
criterion fails.  This negative is not repaired by choosing another prompt.
The supported result is that the complete asynchronous component finishes in
32.3\% of D's complete wall time at this exact point, a reciprocal
3.10$\times$ result\claimref{C2}.  F jointly changes source concurrency from
depth one to depth eight, asynchronous queueing, I/O/compute overlap, and the
complete lifecycle implementation; the comparison does not attribute the
difference independently to \Plan bookkeeping.

\subsection{Integrated Deterministic Faults}
\label{sec:faults}

The Qwen3-Next fault population uses the real graph and demanded objects at
512+8, zero cache, eight O\_DIRECT windows, and a fixed-horizon greedy sampler.
Its deterministic hooks live in a later fault-instrumented binary that retains
the same real adapter/GPU boundary; they are adjacent evidence for the shared
core and are not attributed to the earlier executable used by the 32K+64
experiment.  The earlier executable directly establishes clean no-fault
closure only.
One control and nine deterministic hooks cover short success, EIO,
out-of-order valid completions, late stale completion, wrong object, held
consumer, reset with outstanding source work, shutdown with a live consumer,
and a failed extent inside a bundle.  All ten material jobs are valid.  The
five fatal cells accept zero tokens and logits, publish no invalid readiness or
H2D, explicitly retain the one interrupted \Plan window, and drain the source
executor.  The four recoverable cells retain exact tokens, logits, routes,
source identities, and final cache state against control before ordinary clean
shutdown.

The representative transfer repeats no-fault, short, stale, and held-consumer
cells through the real Gemma 4 graph under a hard 8\GiB cgroup.  All four are
valid.  Short and stale accept zero tokens/logits and begin no invalid H2D;
held-consumer produces bit-identical logits and exact tokens, routes,
materialization count, and final cache occupancy against its material control.
Across all fourteen prespecified Qwen and Gemma cells, swap, max/OOM events,
and foreign compute PIDs are zero.  Every terminal source executor has zero
active tickets and all eight windows free\claimref{C3}.

\begin{table*}[t]
  \centering
  \scriptsize
  \caption{Prespecified fault modes mapped to implementation-independent reuse
  invariants.  Rows group the fourteen cells by transition; they are not
  statistical reliability trials.}
  \label{tab:faultmatrix}
  \begin{tabularx}{\textwidth}{@{}p{3.2cm}p{4.1cm}p{4.5cm}X@{}}
    \toprule
    Mode & Invariant exercised & Required observation & Adapter cells \\
    \midrule
    No fault & Ordinary publication and closure & Exact output; balanced
      tickets, windows, slots, consumers, reset, and shutdown & Qwen, Gemma \\
    Short read; EIO; failed bundle extent & Publish only after complete valid
      movement & Zero accepted outputs, readiness, or invalid H2D; executor
      drains & Qwen; short also Gemma \\
    Stale completion; wrong object & Published version denotes requested
      object and epoch & Mismatch rejected before visibility; zero accepted
      outputs & Qwen; stale also Gemma \\
    Valid out-of-order completion & Legal completion order is not overconstrained
      & Full control-exact output and clean closure & Qwen \\
    Held consumer & No reuse before declared consumers finish & Reuse delayed;
      full control-exact output and balanced release & Qwen, Gemma \\
    Reset with source work; shutdown with live consumer & Drain or invalidate
      outstanding versions & No stale publication; balanced terminal state &
      Qwen \\
    \bottomrule
  \end{tabularx}
\end{table*}

These deterministic cells demonstrate that named invalid transitions are
rejected at the integrated adapter/GPU boundary and that declared legal
reordering remains exact.  They do not sample scheduler interleavings, estimate
failure probability, or prove race freedom, serving isolation, or general
fault tolerance.

\boundedtransfer

\subsection{Resource Admissibility Is Not Product Superiority}

On five software-repository tasks, the aggregate-budget Qwen3-Next path and an
aligned resident Qwen3.6 path both complete all five.  The resident path is
4.965$\times$ faster in total wall time and 14.126$\times$ faster to first useful
action\claimref{C9}.  These are different model paths, not a controlled architecture
comparison, and the ratios do not rank model quality.  They show that storage-
backed sovereignty is valuable when capacity is binding, whereas a fitting
resident model remains preferable when latency is the objective.

\long\def\negativeresults{%
\section{Negative Results and Stopped Lines}
\label{sec:negative}

\begin{table}[t]
  \centering
  \scriptsize
  \caption{Compact negative results.  Each stop is scoped to its prespecified
  identity and acceptance criterion.}
  \label{tab:negative}
  \begin{tabularx}{\columnwidth}{@{}p{0.24\columnwidth}X@{}}
    \toprule
    Line & Result and consequence \\
    \midrule
    GPT-OSS-20B contiguous pack & A local diagnostic removed 82.451\% of its
    direct-service wall, but the physical RTX 3080 Ti cell required 22.146 s,
    removed only 4.014\% of the same direct-service wall target, regressed its control, and
    reached the 6\GiB host cap with 1,349 \texttt{memory.max} events.  The
    observed 1.085\,GB/s carrier would need at least 5.207\,GB/s.  The carrier
    premise did not advance to the next experiment\claimref{C10}. \\
    Llama-3.3-70B IQ2\_M & The published 24,119,299,040-byte artifact was
    reconstructed exactly.  CPU placement of token embedding and all 80 GQA-V
    projections produced a 23,432\MiB GPU peak with 685\MiB margin, but only
    one of four untouched tasks passed.  The 4/4 admission criterion stopped the
    planned 8,192+512 economics cell\claimref{C11}. \\
    \bottomrule
  \end{tabularx}
\end{table}

The GPT-OSS result does not show that packing fails on a faster carrier.  The
Llama result does not establish a general quantization-quality law.  Their role
is procedural: resource or component success is insufficient when the next
prospective criterion fails.

}

\section{Future Work}
\label{sec:prospective}

The matched three-arm source campaign and integrated deterministic-fault
campaign are complete.  A minimally hooked version of the earlier C1 source
would be the direct way to bind transition validation to that executable; a
no-fault cross-build parity cell alone would not establish its fault behavior.
That experiment would strengthen executable-identity integration but is not
required for the explicitly split certificates reported here.  The next
certificate-level transfer question is to apply the same inventory, authority,
equivalence, and lifetime fields to an external storage-offload artifact,
allowing ``insufficient public evidence'' as a valid classification rather than
presuming capability absence.  Guard-mutation tests and systematic schedule
exploration could strengthen causal and interleaving evidence for the lifecycle
protocol.  A same-implementation queue-depth sweep or a simpler asynchronous
direct baseline would be required before attributing the 3.10$\times$ component
result to a narrower mechanism.  A second aggregate-budget model or device
would test breadth.

Broader temporal-amortization and Gemma populations would estimate separate
mechanism generality, not alter the completed principal result.  Until such evidence
exists, the Qwen3.6 timing remains a one-request result and the Gemma H2D and
decode directions remain descriptive matched-\Plan evidence.  No result here
claims a novel direct-I/O primitive, universal policy superiority,
transactional semantics, or general end-to-end fault containment.

\section{Related Work}
\label{sec:related}

\para{General heterogeneous inference}
FlexGen schedules computation and tensors across GPU, CPU, and disk for
throughput-oriented single-GPU generation~\cite{flexgen}.  DeepSpeed Inference
and ZeRO-Inference stream layers from CPU or NVMe, while DeepNVMe supplies
asynchronous host/device tensor I/O and GDS interfaces
\cite{deepspeedinference,zeroinference,deepnvme}.  LLM in a Flash studies
flash-aware execution under limited DRAM~\cite{llminflash}, and FlexInfer uses
direct I/O and fixed tensor windows for on-device inference~\cite{flexinfer}.
Beyond Capacity studies dual host-bypass and GPU-direct flash paths with
measured component latencies and event-driven continuous-batching evaluation
\cite{beyondcapacity}.
These systems establish storage as an inference tier.  This work's residual
claim is a recorded evidence package: explicit representation and semantic-demand
inventories, hard complete-process-tree host charging, audited whole-board GPU
use plus a physical capacity ceiling, full-output-horizon exactness under
$E_{\mathrm{64-out}}$ at the principal Qwen3-Next experiment, a distinct matched ownership
and complete asynchronous-component comparison, and named later-build
cross-adapter fail-closed transitions.

\para{Sparse expert loading and caching}
MoE-Infinity uses sparsity-aware expert caching on personal machines
\cite{moeinfinity}.  SpecOffload uses speculative decoding to unlock otherwise
idle GPU capacity in an offloaded engine~\cite{specoffload}; FlashMoE studies
learned cache replacement~\cite{flashmoe}; and FluxMoE decouples expert
residency using stable logical addresses~\cite{fluxmoe}.  Recent cache-ownership
work demonstrates hard cgroup capacity for expert replay and argues that kernel
LRU can be competitive with user-space ownership~\cite{expertcacheownership}.
WiSP frames MoE serving as working-set management and reports byte-identical
outputs and up to 1.95$\times$ decode throughput over static offload at matched
GPU memory~\cite{wisppaper}.
These results prevent cgroups, caching, or stable addresses from being treated
as standalone novelty.

\para{Closest engineering disclosures}
MoE-Direct combines byte-preserving NVMe reads, bounded slots, generations,
pins, and scoped token identity in a public preview artifact
\cite{moedirect}.  Its source, preview binary, and strongest reported execution
identities do not fully coincide, and complete RAM headroom is not reported.
MESH combines O\_DIRECT/io\_uring, fixed expert slots, reader counts, and
policies, but at the evidence cutoff it is an unmerged KTransformers pull
request rather than mainline behavior or a peer-reviewed result~\cite{mesh}.
Palm's SSD path is merged public engineering code without an associated paper;
its large-model path is therefore cited as an artifact, not paper evidence
\cite{palm}.  WISP independently combines VRAM/RAM/NVMe tiers, double-buffered
asynchronous prefetch, RAM watermarks, and VRAM planning in a public engine
\cite{wispengine}.  Colibr\`i is the closest newly audited artifact in mechanism
breadth: its fast-moving multi-family engine includes RAM/NVMe/VRAM tiers,
O\_DIRECT, optional io\_uring, bounded asynchronous loading, byte-preserving
formats, token-exact harnesses, resource planning, and documented cgroup
operation~\cite{colibri}.  The WiSP preprint and WISP engine are unrelated
despite the name collision.

The refreshed audit covers public records through 20 August 2026.  It makes
novelty based on direct I/O, asynchronous streaming, three-tier placement,
exact byte formats, or cgroups alone untenable.  At the audited identities, no
single disclosure was found to package the complete set of principal claims as
one recorded evidence package: demand and representation arithmetic with graded authorities,
the six-pair byte-matched P/D/F experiment, and all fourteen prespecified
real-adapter transition cells.  This is a cutoff-scoped public-record finding,
not proof that another system lacks the underlying capability.  Our evaluation likewise
discloses that its C1 execution and C3 transition evidence use different
binaries.  The evidence package, rather than mechanism composition or a single-
executable conjunction, is the residual claim.

\begin{table*}[t]
  \centering
  \scriptsize
  \caption{Closest-record audit through 20 August 2026.  Entries describe the
  cited public identity, not latent system capability.  ``Partial'' means that
  some evidence exists but not the complete certificate field used here.}
  \label{tab:relatedmatrix}
  \begin{tabularx}{\textwidth}{@{}p{2.3cm}p{2.7cm}p{3.0cm}p{2.7cm}X@{}}
    \toprule
    Disclosure & Inventory / demand & Resource authority & Exactness horizon &
      Matched ownership / integrated faults \\
    \midrule
    This work & Exact $T_{\mathrm{repr}}$ and $U_{\mathrm{sem}}^{\mathrm{LB}}$
      & Host hard; GPU audited; physical board ceiling & All 64 complete logits,
      tokens, response, routes, and identities & Six-pair P/D/F; fourteen named
      later-build cells across two adapters \\
    WiSP~\cite{wisppaper} & Working-set model; partial certificate mapping &
      Matched GPU memory; partial aggregate authority & Byte-identical outputs
      at reported scope & No corresponding fixed-identity pair in the audited paper \\
    WISP~\cite{wispengine} & VRAM/RAM/NVMe planning; partial & RAM watermark and
      VRAM planning; partial & Verified operating point; no immutable paper
      horizon & No corresponding fixed-identity campaign in the audited artifact \\
    Colibr\`i~\cite{colibri} & Multi-family byte-preserving formats; partial &
      Resource planning and cgroup guidance; partial & Token-exact harnesses;
      partial & No corresponding fixed-identity campaign in the audited revision \\
    MoE-Direct / MESH~\cite{moedirect,mesh} & Extents, slots, and versioning;
      partial & Incomplete coincident authority at audited identities & Scoped
      token or implementation evidence & No corresponding cross-adapter fault
      population in audited identities \\
    \bottomrule
  \end{tabularx}
\end{table*}

\section{Limitations}
\label{sec:limitations}

The principal experiment covers one Qwen3-Next quantization, mechanics prompt, tested
runtime, RTX 3090, and 64-output horizon.  Its semantic-demand quantity is a
lower bound over managed expert bundles; it does not claim that all demanded
objects are simultaneously live or inventory every nonexpert tensor touched by
the graph.  The full-output oracle does not expose recurrent/hybrid-state bytes
or establish equality to an unmodified upstream runtime.
The matched source result adds six paired observations at one natural workload;
it does not establish a workload population or separate concurrency, overlap,
and lifecycle implementation.  P is one buffered implementation, not a theorem
about every bounded page-cache design.

The RTX 3080 Ti measurement record reports the GPU, driver, virtual source device, and
resource observations but not the physical CPU or provider storage model.  Its
timings therefore cannot support a hardware-normalized storage claim.

Complete process-tree charging and whole-board observation close the largest
hidden-residency channels, but GPU use is audited rather than hard-partitioned.
Here process-tree charging means cgroup-attributed memory, not every host byte
causally associated with execution; per-class \texttt{memory.stat} and separate
driver-pinned attribution were not archived.  This boundary does not weaken
the independent semantic-demand nonresidency result against Linux-visible host
capacity plus the physical board, but it limits the 11\GiB resource predicate
to the declared cgroup-charge contract.
Opaque firmware accounting remains bounded by the vendor interface.  The
Gemma representative fault campaign does not assert page-cache first-charger
ownership for its nonexpert load.  Deterministic integrated fault injection
does not prove race freedom, stochastic reliability, or serving isolation, and
the Qwen hooks use a later fault-instrumented binary rather than the earlier
principal-experiment
executable.  Cross-adapter evidence is limited to Qwen and Gemma.  The thin MTP
margins do not establish robust deployment headroom.  Finally, storage-backed
execution is a capacity method; it can be substantially slower than a smaller
resident model.

\section{Conclusion}

Memory capacity is a credible systems claim only when its budget, canonical
bytes, execution oracle, and asynchronous lifetimes are explicit.  The architecture
separates storage ownership, physical residency, stable execution addresses,
and scheduling.  Its C1 certificate binds a 45.08\GiB native
representation, a 43.59375\GiB semantic-demand lower bound beyond the declared
and host-hard/device-physical envelopes, full-output-horizon exactness under
$E_{\mathrm{64-out}}$, hard complete-process-tree cgroup-charge accounting, audited
whole-board GPU margin, and clean no-fault closure.  Together these establish
resource admissibility under the prespecified hard-margin contract\claimref{C1}.

Relative to the tested buffered P path, direct source ownership changes the
prespecified admission outcome from exact-but-at-limit to exact with positive
margin and zero limit events.  Six matched direct-source pairs show that the
complete eight-window asynchronous component finishes in 32.3\% of the
blocking one-window reader's complete wall time at identical physical bytes per
output\claimref{C2}.  In separate later fault-instrumented binaries, all fourteen prespecified
deterministic cells support fail-closed behavior for named transitions across
real Qwen3-Next and Gemma 4 adapter/GPU paths\claimref{C3}.  The appendix preserves exact temporal
amortization, Qwen-specific prefetch, the shared-\Plan comparison and its
prespecified F1 miss,
Gemma transfer, and stopped lines at their original evidence identities.

The failed churn premise, joint concurrency/bookkeeping contrast, deterministic
fault schedule, audited rather than hard-partitioned GPU budget, and narrow
model/hardware identities bound the claim.  They forbid independent \Plan
causality, race-freedom, serving, or general fault-tolerance language.  A
fault-instrumented build from the earlier source, a broader model/device
principal experiment, an independent
upstream oracle, and broader schedule exploration remain legitimate ways to
increase identity integration, breadth, or independence; none is silently
promoted into the scoped result.

\appendix
\additionalexperiments
\negativeresults

\section{Paper-to-Artifact Claim Ledger}
\label{sec:claimledger}

Every superscript \textsf{C\#} resolves to one entry in
\texttt{claims/claim-ledger.json}.  In the accompanying artifact,
\texttt{./artifact verify --claim C\#} verifies source digests, locators, and
expected assertions; \texttt{./artifact replay --claim C\#} recomputes the
packaged summary where replay is applicable.  Each entry contains the paper
statement, immutable result path, full SHA-256 digest, JSON pointers or a text
locator, and both commands.  Table~\ref{tab:claimledger} abbreviates the digests
for print; the machine-readable ledger preserves all 64 hexadecimal digits.
Artifact release \texttt{v0.1.2}: \url{https://github.com/gustavgauge/memory-sovereign-inference-artifact/releases/tag/v0.1.2}.

\begin{table*}[t]
  \centering
  \scriptsize
  \caption{Mechanical paper-to-artifact claim map.  The stable artifact commands
  substitute the listed ID for \texttt{C\#}.}
  \label{tab:claimledger}
  \begin{tabularx}{\textwidth}{@{}p{0.75cm}p{5.2cm}p{5.4cm}X@{}}
    \toprule
    ID & Headline family & Immutable compact result & SHA-256 prefix \\
    \midrule
    \hypertarget{claim:C1}{C1} & Semantic demand beyond full residency & Publication claim record & \texttt{af894187b754} \\
    \hypertarget{claim:C2}{C2} & Six-pair complete asynchronous component & Publication claim record & \texttt{1fad89aa4103} \\
    \hypertarget{claim:C3}{C3} & All fourteen prespecified fault/control cells & Publication claim record & \texttt{c2ca7498e92c} \\
    \hypertarget{claim:C4}{C4} & Qwen3.6 MTP temporal amortization & Publication claim record & \texttt{3aa1cd4f8545} \\
    \hypertarget{claim:C5}{C5} & Qwen-specific natural utility & Publication claim record & \texttt{67922edcd38e} \\
    \hypertarget{claim:C6}{C6} & Shared-\Plan comparison and preserved F1 miss & Publication claim record & \texttt{72b59f174b88} \\
    \hypertarget{claim:C7}{C7} & Gemma cache transfer & Publication claim record & \texttt{ef3215cfd764} \\
    \hypertarget{claim:C8}{C8} & Gemma bounded direct source & Publication claim record & \texttt{4432f76df666} \\
    \hypertarget{claim:C9}{C9} & Cross-model product comparison & Publication claim record & \texttt{b3a576f4c5a8} \\
    \hypertarget{claim:C10}{C10} & GPT-OSS storage-device stop & Publication claim record & \texttt{03a01c6ededb} \\
    \hypertarget{claim:C11}{C11} & Llama task-admission stop & Publication claim record & \texttt{cc4865e51d59} \\
    \hypertarget{claim:C12}{C12} & Shared-\Plan parity process & Publication claim record & \texttt{88e7fa900277} \\
    \bottomrule
  \end{tabularx}
\end{table*}

\section{Exact Claim Identities}
\label{sec:identities}

The following SHA-256 values are the exact model or model-inventory identities
used by the claim-bearing campaigns.  Line breaks are typographic only.

\noindent\begin{minipage}{\columnwidth}
\textbf{Qwen3.6 target model}\par
\fullhash{f0863123}{e9f79c1f}{43bb6d4f}{2d7bc645}{41a4e92f}{7242c3dd}{ae649731}{c03cdd09}
\end{minipage}\par\smallskip
\noindent\begin{minipage}{\columnwidth}
\textbf{Qwen3.6 target+MTP composite}\par
\fullhash{79e80b7a}{2237412a}{ad13bdd4}{0151ab0c}{04a10442}{9182f9b7}{19f0bb47}{d181b7cd}
\end{minipage}\par\smallskip
\noindent\begin{minipage}{\columnwidth}
\textbf{Qwen3-Coder AWQ checkpoint digest list}\par
\fullhash{9da173bf}{d3be51db}{5c781ddd}{cdf3d05b}{7fa45ddd}{08abe7f0}{dc323c49}{8fe5f218}
\end{minipage}\par\smallskip
\noindent\begin{minipage}{\columnwidth}
\textbf{Qwen3-Coder prepacked store}\par
\fullhash{8a89a889}{0d2e9d08}{bf7d74d8}{71b1ccc4}{a7a19d74}{ceb944b0}{0b237d09}{39dffc11}
\end{minipage}\par\smallskip
\noindent\begin{minipage}{\columnwidth}
\textbf{Gemma 4 QAT Q4\_0 model}\par
\fullhash{3eca3b8f}{6d7baf21}{8a7dd6bb}{a5fb59a5}{6ee25fe2}{d567b6f5}{f589b4f6}{97eca51d}
\end{minipage}\par\smallskip
\noindent\begin{minipage}{\columnwidth}
\textbf{Qwen3-Next Q4\_K\_M model file}\par
\fullhash{d103b273}{3ec1012a}{52d01edd}{a66b7e5c}{24ae5050}{8c9f99f5}{297ea459}{ef3c061a}
\end{minipage}

\smallskip\noindent\textbf{Runtime commits.}
\begin{description}[leftmargin=0pt,labelindent=0pt,style=nextline,itemsep=2pt]
  \item[Qwen3.6 base]
  \commitid{1dfb398e}{a4317c92}{04e40aa8}{60c0c251}{ab74d8a4}
  \item[Qwen AWQ]
  \commitid{b29d0006}{7fe3d641}{ba2995b7}{c693d8be}{a8c83a8f}
  \item[Gemma base / adapter]
  \commitid{40d5358d}{3c730b81}{729ba81c}{d5c44ed5}{96d02510}\quad
  \commitid{05f332a0}{b031aa18}{041622e2}{6f37e6a7}{c1fe82f0}
  \item[Qwen3-Next base / adapter]
  \commitid{8ef78e64}{4f559db4}{e8716b59}{bf76b8e1}{1619337d}\quad
  \commitid{72066050}{99424f7a}{d7ca78c2}{783a60ec}{60f810f4}
\end{description}

\bibliographystyle{plain}
\bibliography{references}

\end{document}